\documentclass[twocolumn]{revtex4-1}
\usepackage[T1]{fontenc}
\usepackage[latin1]{inputenc}
\usepackage{amsmath}
\usepackage{amssymb}
\usepackage{graphicx}
\usepackage{color}
\usepackage[unicode=true,pdfusetitle,
 bookmarks=true,bookmarksnumbered=false,bookmarksopen=false,
 breaklinks=false,pdfborder={0 0 1},backref=false,colorlinks=false]
 {hyperref}

\makeatletter

\usepackage{dcolumn}
\usepackage{bm}

\usepackage{latexsym}
\usepackage{amsthm}\usepackage{verbatim}
\usepackage{vmargin}
\usepackage{url}
\usepackage{array}

\graphicspath{{./Figures_articles_bassins_v10/}}

\makeatother

\begin{document}

\hspace{5.2in} 
\title{Equalization of pulse timings in an excitable microlaser system with delay}
\author{Soizic Terrien$^{*}$}
\affiliation{The Dodd-Walls Centre for Photonic and Quantum Technologies, The University
of Auckland, New Zealand}
\author{V. Anirudh Pammi$^{*}$}
\affiliation{Centre de Nanosciences et de Nanotechnologies, C2N-UMR9001, CNRS,
Universit\'e Paris-Sud, Universit\'e Paris-Saclay, Site de Marcoussis,
Route de Nozay, 91460 Marcoussis, France}
\author{Neil G. R. Broderick}
\affiliation{The Dodd-Walls Centre for Photonic and Quantum Technologies, The University
of Auckland, New Zealand}
\author{R\'emy Braive}
\affiliation{Centre de Nanosciences et de Nanotechnologies, C2N-UMR9001, CNRS,
Universit\'e Paris-Sud, Universit\'e Paris-Saclay, Site de Marcoussis,
Route de Nozay, 91460 Marcoussis, France}
\author{Gr\'egoire Beaudoin}
\affiliation{Centre de Nanosciences et de Nanotechnologies, C2N-UMR9001, CNRS,
Universit\'e Paris-Sud, Universit\'e Paris-Saclay, Site de Marcoussis,
Route de Nozay, 91460 Marcoussis, France}
\author{Isabelle Sagnes}
\affiliation{Centre de Nanosciences et de Nanotechnologies, C2N-UMR9001, CNRS,
Universit\'e Paris-Sud, Universit\'e Paris-Saclay, Site de Marcoussis,
Route de Nozay, 91460 Marcoussis, France}
\author{Bernd Krauskopf}
\affiliation{The Dodd-Walls Centre for Photonic and Quantum Technologies, The University
of Auckland, New Zealand}
\author{Sylvain Barbay}
\affiliation{Centre de Nanosciences et de Nanotechnologies, C2N-UMR9001, CNRS,
Universit\'e Paris-Sud, Universit\'e Paris-Saclay, Site de Marcoussis,
Route de Nozay, 91460 Marcoussis, France}

\begin{abstract}
An excitable semiconductor micropillar laser with delayed optical
feedback is able to regenerate pulses by 
the excitable response of the laser. It has been shown that 
almost any pulse sequence can, in principle,
be excited and regenerated by this system over short periods of time. We show experimentally
and numerically that this is not true anymore in the long
term: rather, the
system settles down to a stable 
periodic orbit with equalized timing between pulses. Several such
attracting periodic regimes with different numbers of equalized pulse timing
may coexist and we study how they can be accessed with
single external optical pulses of sufficient strength that need to be
timed appropriately.
Since the observed timing equalization and switching characteristics
are generated by excitability in combination with delayed feedback,
our results will be of relevance beyond the particular case of photonics,
especially in neuroscience.
\end{abstract}

\maketitle

Excitability is observed in many natural and artificial systems, including
spiking neurons, cardiac cells, and semiconductor lasers. It corresponds
to the all-or-none response in the form of a single spike to an input
external perturbation, depending on whether or not the amplitude of
the perturbation exceeds the so-called excitable threshold \citep{IzhikevichBook}.
When subject to delayed feedback, an excitable system can either
remain in its quiet state for small external perturbations
or, with an adequate control pulse of sufficient strength, it can then
regenerate its own excitable response after the reinjection time $\tau$.
This very general mechanism for self-pulsations has been 
implemented
in different optical systems, including a coherently driven VCSEL
\citep{GarbinNC15}, a VCSEL subject to opto-electronic feedback \citep{marino2017chaos},
coupled semiconductor lasers \citep{KelleherPRE10}, a photonic resonator
with optical self-feedback \citep{RomeiraNSR16} and a micropillar
laser with integrated saturable absorber \citep{TerrienPRA17}.

Since almost arbitrary pulse timing patterns can, in
  principle, be excited and regenerated after each delay, regenerative
  dynamics can be of particular 
interest for producing complex optically-controllable temporal pulsing patterns
\citep{MarconiPRL14,JangNatCom2015,camelin2016electrical,JavaloyesPRL16,Terrien2018OL} 
or for spike-based optical memory
applications \citep{GarbinNC15,RomeiraNSR16,ShastriSR16,Terrien2018OL}. 
In the context of biological spiking neurons, delayed self-connections
have also been recognized to play a central role 
in the persistent regeneration of input stimuli \citep{FossPRL96,ConnorsN02,ChaudhuriNN16}.
Systems with delay generally display rich dynamics with coexistence between different
types of attractors \citep{IkedaPRL82,YanchukPRE09}. Consequently, it
is an important question to determine the long-term dynamics 
of regenerative pulsing in excitable systems with delay. 

Here we show experimentally and theoretically that it is not possible
to regenerate arbitrary timing patterns in the long term: any
triggered pulse pattern will equalize, upon sufficiently
many successive
regenerations, to an equidistant pulse train. Hence, positional information of
non-equalized pulse patterns is preserved only for short periods of time
and cannot be sustained in the long-term. 
Since the long-term information is encoded in the number of pulses in
the feedback loop, we also investigate how one can switch between
different equalized stable pulse trains. From a theoretical
perspective, this is related to the structure of their basins of
attraction, which we investigate numerically. The underlying physics of equalization as well as of switching between patterns 
is entirely the result of an interplay between the timescale of the slow dynamical variable (here the net gain dynamics \citep{NizettePDNP06}) 
and the latency time of the excitable system \citep{SelmiPRE16}. As
such, this mechanism is very general for excitable systems subject to
delayed feedback.

  In this Letter, we consider an excitable microlaser with integrated
saturable absorber \citep{BarbayAPL05,ElsassEPJD10,BarbayOL11,SelmiPRL14}
and delayed optical feedback \citep{TerrienPRA17,Terrien2018OL}.
Thanks to its small footprint, sub-nanosecond response time and easy
bi-dimensional integration, this device is of particular interest
for applications ranging from photonic spike processing
\citep{NahmiasIJoSTiQE13,PrucnalAiOaP16} 
for efficient optical communications applications to ultrafast artificial
neural networks. Without feedback, the solitary micropillar laser
is excitable in a wide pump parameter region below the self-pulsing
threshold \citep{BarbayOL11} and displays various neuromimetic properties
such as a relative refractory period \citep{SelmiPRL14}, temporal
summation \citep{SelmiOL15} and spike latency \citep{SelmiPRE16,ErneuxPRE18}.
In the presence of delayed optical feedback, it sustains trains of regenerative optical
pulses, which can be asymmetrically perturbed by noise
\citep{TerrienPRA17} or added and erased by single optical
perturbations \citep{Terrien2018OL}.

The experimental setup consists of a micropillar laser with two gain
and one saturable absorber quantum wells, which emits light at a wavelength
close to 980 nm. Part of the output light is sent back into the micropillar
after a delay $\tau$, through free-space propagation and reflection
by a mirror at several tens of cm from the laser. A beamsplitter
 in the optical feedback path (R/T=70/30) redirects some
of the light to a fast avalanche photodetector or a camera. The micropillar
is pumped at 800nm, and short optical perturbations of 80 ps duration
can be sent by a mode-locked Ti:Sa laser emitting at the pump wavelength.

\begin{figure}[!t]
\includegraphics[width=1\columnwidth]{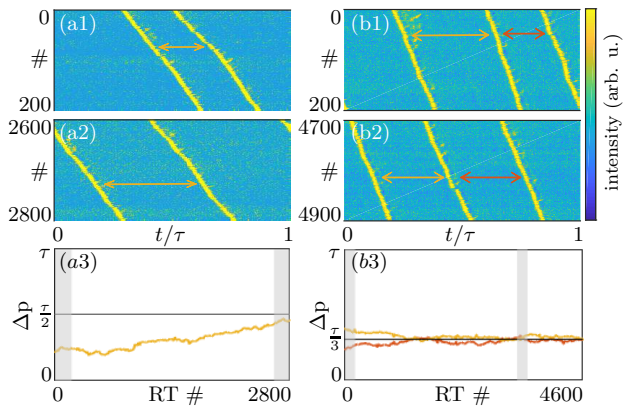} \caption{(a),(b): Experimental pulse trains following two (a) and three (b)
perturbations, for a feedback delay of 8.2ns. Shown are the pseudo-space
representation of the time traces observed shortly after the perturbations
[(a1) and (b1)] and after several thousands of round trips [(a2) and
(b2)]. The pulse-to-pulse timing $\Delta p$ (as shown by arrows in
(a1) and (b1)) is plotted versus the round trip number in panels (a3)
and (b3), where the shaded areas are the time segments represented
in (a1--a2) and (b1--b2).}
\label{fig:expe_convergence}
\end{figure}

The experimental system is modeled accurately by the Yamada rate equations
with incoherent delayed feedback
\citep{Yamada93,KrauskopfWalker,BarbayOL11,SelmiPRL14,TerrienPRA17,Terrien2018OL} ---
a system of three DDEs for the dimensionless gain $G$, absorption
$Q$ and intensity $I$: 
\begin{equation}
\begin{split}\dot{G} & =\gamma_{G}(A-G-GI),\\
\dot{Q} & =\gamma_{Q}(B-Q-aQI),\\
\dot{I} & =(G-Q-1)I+\kappa I(t-\tau).
\end{split}
\label{eq:yam}
\end{equation}
Here, $A$ is the scaled pump parameter (relative to pump at transparency),
$B$ is the non-saturable absorption, $a$ is the saturation parameter,
and $\gamma_{G}$ and $\gamma_{Q}$ are the carrier recombination
rates in the gain and absorber media, respectively. The optical feedback
is described by the delayed term in the intensity equation, where
$\kappa$ is the feedback strength and $\tau$ is the feedback delay.
We consider here the same parameter values as in \citep{Terrien2018OL}:
$A=2.4$, $B=2.2$, $\gamma_{G}=0.01$, $\gamma_{Q}=0.02$, $a=5$,
$\kappa=0.05$, $\tau=1100$. These are chosen both to match the known
physical parameters and the experimental observations. In particular,
the small values of $\gamma_{G}$ and $\gamma_{Q}$ account for the
slow non-radiative recombination of the carriers in the gain and absorber
media, compared to the fast timescale of the laser field intensity.

Figure \ref{fig:expe_convergence}(a) and (b) show experimental results
on the convergence of irregularly spaced pulse trains to regularly
spaced ones following two and three external perturbations, respectively.
In panels (a1)--(a2) and (b1)--(b2) the temporal traces are folded at
approximately the delay $\tau$ and stacked vertically in a pseudo-space
representation \citep{giacomelli1998multiple}. Initially non-equidistant
pulse trains in the external cavity become equidistant after several
thousands of round trips, as shown in panels (a2) and (b2). The slow
convergence towards equidistant pulsing patterns is highlighted in panels
(a3) and (b3), which represent the pulse-to-pulse timing $\Delta p$
versus the round trip number, from the instant when the pulse trains
are triggered by external perturbations. The pulse-to-pulse timing
$\Delta p$ slowly converges to a value close to a half or a third
of the delay time $\tau$, respectively, as equidistant pulsing is
approached. This slow
convergence rate is of the order of a few ps per round trip, to be
compared to the pulse duration of approximately 200 ps. It can be observed
in the experiment only over long time periods.
In contrast to the ultra weak soliton interaction observed in \citep{jang2013ultraweak},
it can be simply explained by the variation of the response latency
time of the excitable microlaser in the slowly recovering landscape
of the net gain dynamics \citep{DubbeldamPRE99,SelmiPRE16,ErneuxPRE18,Terrien2018OL}.
This response latency becomes identical only when all the re-injected
pulses experience an identical net gain \citep{SelmiPRL14,Terrien2018OL}.
This configuration corresponds to a stable equidistant pulse train
in the case of a fast SA. The stochastic fluctuations of the pulse-to-pulse
timing are explained by the presence of pump noise in the system,
which induces stochastic fluctuations of the microlaser net gain \citep{TerrienPRA17}.

\begin{figure}[t!]
\includegraphics[width=1\linewidth]{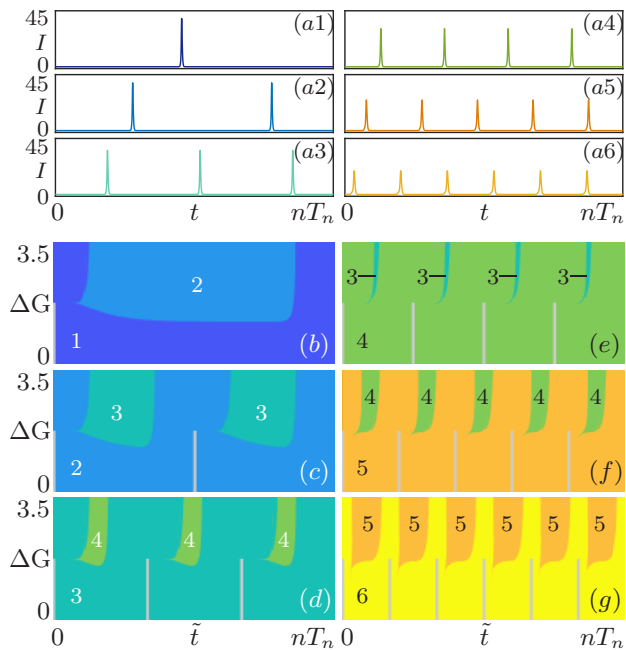} \caption{(a1)--(a6): Intensity time series of the stable periodic pulsing regimes of equations (\ref{eq:yam}) represented over $n$ periods $T$, with $n$ the number of pulses in the span of the delay $\tau$. (b)--(g): Basins of attraction of (\ref{eq:yam}) in the plane of timing $\tilde{t}$
and amplitude $\Delta G$ of a gain perturbation, when one (b) to six (g) equidistant pulses initially exist in the external cavity. The color represents the number of pulses observed in the long-term in the external cavity (see (a1)--(a6) for the color code), and the vertical gray lines indicate the timing of the pre-existing pulses.}
\label{fig:basins_dG}
\end{figure}

The Yamada model with feedback (\ref{eq:yam}) shows excellent
agreement with these experimental 
observations. Its phase portraits are calculated with the continuation
toolbox DDE-Biftool \citep{Engelborghs_01,Sieber_biftool} and show
a high degree of multistability. In particular, one stable equilibrium
corresponding to the non-lasing solution coexists with six stable
periodic solutions with different periods $T_n$ and equalized pulse
timings, whose corresponding time 
series are represented in figure \ref{fig:basins_dG}(a1)--(a6). Their
periods are close to sub-multiples of the delay time 
$\tau$ \citep{TerrienSIADS17}, and they are hereafter referred to
as 1-pulse solution, 2-pulse solution, and so on. A Floquet
stability analysis has confirmed that the solutions with two to six
coexisting pulses in the external cavity are only weakly stable \citep{Terrien2018OL}.
Importantly, for the chosen parameters, no stable solution exists that
corresponds to pulse trains with non-equidistant pulses. Therefore,
all pulsing dynamics must converge towards one of the attracting
periodic solution represented in Figure \ref{fig:basins_dG}(a). As
observed in the experiment, the convergence to the weakly stable pulsing
regimes occurs on a slow timescale \citep{Terrien2018OL} compared
to the feedback delay time.

The final state of a multistable system depends crucially on the initial
conditions. For each attractor in figure \ref{fig:basins_dG}(a),
its basin of attraction is the set of initial conditions for which
the system settles on that attractor after transient dynamics. 
From a practical point of view, the structure of these basins of
attraction gives essential information on how to access the different
coexisting stable pulsing regimes and, as such, on how to control the
long-term dynamics of the multistable system \citep{PisarchikPR14}.
In systems of ordinary differential
equations (ODEs) with up to three dimensions, the invariant manifolds
that bound the different basins of attraction can be calculated with
advanced numerical methods \citep{Krauskopf2005IJBC}.
However, we deal here with a system of delay-differential equations
(DDEs), whose phase space is intrinsically of infinite dimension (see
e.g. \citep{balachandran2009delay}). The numerical continuation of
the projections of such invariant manifolds is, hence, more complex
\citep{krauskopf2003computing,KeaneNL18}, and we rather 
integrate (\ref{eq:yam}) numerically to map the basins of attraction.

Figure \ref{fig:basins_dG}(b)--(g) represents the long-term
effect of an additive perturbation on the gain variable
$G$ of amplitude $\Delta G$, when it is given at a relative timing $\tilde t$ 
in a stable equidistant pulsing regime of (\ref{eq:yam}), where
$\tilde t =0$ is the reinjection time of a pre-existing pulse in the
microlaser. The color code 
represents the attractor on which the system settles in the long term
(\emph{i.e} after the transient dynamics). When the system is initially
in the $n$-pulse regime with $n=1,2$ and $3$ [panels (b)--(d)],
the perturbation triggers an additional pulse and the system can settle
to the ($n$+1)-pulse regime, for suitable amplitude and timing
of the perturbation. In Figure \ref{fig:basins_dG}(b)--(g) we first
observe that there is a minimum perturbation amplitude ($\Delta G_{min}\simeq1.5$)
to induce a change in the pulsing regime. For $\Delta G>\Delta G_{min}$,
a perturbation has no effect on the overall number of coexisting pulses
if it is introduced immediately before 
or immediately after a pre-existing pulse
is reinjected into the microlaser. In the first case, a new sustained
pulse train is triggered, but its refractory period prevents the pre-existing
pulse train from being regenerated, thus resulting in a global \emph{retiming}
of the pulse train \citep{Terrien2018OL}. In the second case, the
perturbation is introduced in the refractory period of a pre-existing
pulse, and the gain in the micropillar laser is not sufficiently high
for a new pulse to be sustained. Note that the effect of the relative
refractory period \citep{SelmiPRL14} is clearly visible in the initial
negative slopes of the bottom left boundaries of the new stable pulsing regimes. 
When the perturbation is introduced away from
the previous zones, a new sustained pulse train is triggered. Figure \ref{fig:basins_dG}(b)--(d) 
shows that $\Delta G_{min}$ globally increases with the number $n$
of the initial $n$-pulse regime, while the time window to trigger an
additional sustained pulse and switch to the ($n+1$)-pulse regime shrinks.

When the initial stable regime is the $n$-pulse regime
with $n>3$, figure \ref{fig:basins_dG}(e)--(g) shows that, as before,
a perturbation has no effect on the long-term dynamics if it is introduced
in a (generally small) time window around a pre-existing pulse. However, and in
contrast to the previous cases, it is no longer possible to reach the
($n$+1)-pulse regime. A perturbation with appropriate timing
and amplitude now only brings the system to the ($n-1$)-pulse regime,
thus removing one pulse from a pre-exisiting pulse train. For this
to happen, the time window in which the perturbation has to be introduced
widens with increasing $n$. It is thus more likely, e.g., to take
the system away from the 6-pulse regime than from the 4-pulse
regime. Although regimes with more than four coexisting pulses
in the external cavity exist and are stable, figure \ref{fig:basins_dG}(e)--(g) clearly suggests
that they could be particularly difficult to observe in practice with
this perturbation method, and this is confirmed by the experiment.

\begin{figure}[t!]
\includegraphics[width=1\linewidth]{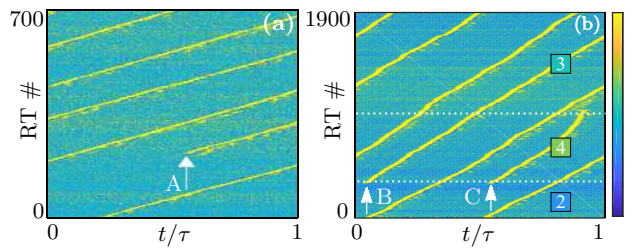}
\caption{Pseudo-space representation of experimental time traces, with a feedback
delay of 8.2ns, showing the effect of external perturbations (indicated
by arrows) of different timings, for a 1-pulse (a) and a 2-pulse (b)
regime.}
\label{fig:expe_timing}
\end{figure}

In the experiment, the ability of an optical perturbation to trigger
a second and a third sustained pulse trains has been shown in figure
\ref{fig:expe_convergence}. Figure \ref{fig:expe_timing} highlights
the influence of the perturbation timing on the long-term dynamics
of the microlaser with delayed optical feedback. Panel (a) shows that
a perturbation (labelled A) introduced far from a pre-existing pulse train can
make the system switch to the 2-pulse regime, in excellent qualitative
agreement with the theoretical results of figure \ref{fig:basins_dG}(a).
Had the perturbation been sent closer to an existing pulse, it would
have either globally retimed the initial pulse train, leaving the
system in the same state, if introduced slightly earlier; if introduced
slightly later than the regenerated pulse, it would have had no effect
since it would have fallen in the refractory period of the pre-existing
pulse. The net effect in the two cases is the same, as far as
the asymptotic state is concerned.

\begin{figure}[t!]
\includegraphics[width=1\linewidth]{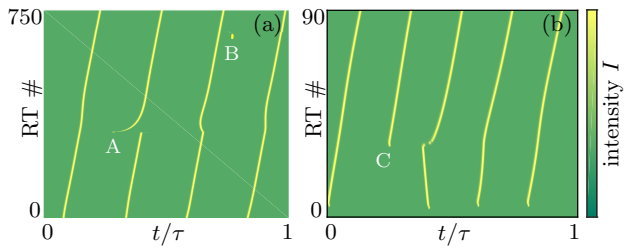} \caption{Pseudo-space
  representation of simulated time traces, showing the 
effect of external perturbations. In (a), starting from the stable 4-pulse
regime, two perturbations are introduced before (A) and after (B) a
reinjected pulse, leaving the system in the stable 4-pulse regime.
In (b), the 5-pulse regime is approached from a transient regime
with four pulses per round trip.}
\label{fig:theo_timing}
\end{figure}

Figure \ref{fig:expe_timing}(b) illustrates the influence of perturbations
with different timings when two and three pulse trains pre-exist in
the feedback cavity: starting from the 2-pulse regime, a third sustained
pulse train is triggered by an optical perturbation (labelled B). The fourth perturbation
(labelled C) is introduced shorty afterwards with a similar relative timing with respect to
the two pre-existing pulses (\emph{i.e} in the pseudo-space representation
it appears to be half-way in between two pre-existing pulses). However,
it only triggers a transient pulse; hence, it does not affect the long-term
dynamics of the system which settles back on the 3-pulse solution
after a few hundreds of round trips. As predicted by the theory in
figure \ref{fig:basins_dG}, these results confirm that triggering
new sustained pulse trains becomes more and more challenging when
the number of pre-existing pulse trains in the external cavity increases.
In particular, figure \ref{fig:basins_dG}(e) shows that an external
perturbation cannot trigger a fifth sustained pulse train in the
model when the system is in the 4-pulse regime. The temporal traces
associated to this case are plotted in \ref{fig:theo_timing}(a).
As observed in the experiment, perturbations sent slightly before
or after a pulse have no effect on the long term dynamics and leave
the system in the same state. By contrast, figure \ref{fig:theo_timing}(b)
demonstrates that it is nevertheless possible to reach the 5-pulse
solution by sending the fifth perturbation (labelled C) during the transient dynamics,
when four pre-existing pulses are still far from their stable
configuration. In general, all the stable $n$-pulse regimes can be
accessed from suitable transient dynamics by addition of single or multiple
perturbation pulses.
Overall, the experiment and the numerical analysis show excellent
qualitative agreement in terms of the influence of the perturbation
timing on the long-term dynamics of the system.

We point out that, in the experiment, external perturbations can be
sent either coherently (at the laser emission wavelength) or incoherently
(at the pump wavelength) \citep{SelmiPRE16}, which corresponds in
system (\ref{eq:yam}) to perturbations on the intensity variable
$I$ or on the gain variable $G$, respectively. The basins were also
mapped with coherent perturbations $\Delta I$ on the intensity variable $I$.
Apart from differences observed mainly in the finer details of the
basin boundaries, which is related to intersections of higher dimension
manifolds, the structure of the basins of attraction is qualitatively
as those shown in figure \ref{fig:basins_dG}(b)--(g). Interestingly, this strongly suggests
that what does matter is the strength and timing of the perturbation,
rather than the exact way the perturbation is introduced.

In conclusion, we have shown how any initial pulsing pattern
equalizes to an equidistant pulse train in the excitable micropillar
laser with delayed optical feedback. Different stable equalized periodic
orbits with different number of pulses in the feedback loop 
are sustained, and they can be accessed by means of single
optical input pulses. Our study of the basins of attraction has shown
how that, depending on the timing, a new pulse can be added when the
number of initial equalized pulses is low, or a pulse can be
subtracted from a sequence when the
number of initial equalized pulses is larger. The experimental
and theoretical results are in good agreement and allow a clear interpretation
of the observed physical phenomena, which are based on small pulse-to-pulse
differences generated by the slow carrier dynamics of the gain and
absorber media. They provide a 
global physical picture of the short-term and long-term 
dynamics of regenerative pulse coexistence. 

In terms of memory applications, any input pulse
pattern will necessarily converge to one of the sustained and stable equalized
pulse trains. While the information encoded in non-equal pulse spacing
will be lost in the medium to long term, this device has the ability
to converge to a given number of pulses in the feedback loop from an
imperfect input \citep{Hertz18}. This approximation property is linked to the
fine structure of the infinite-dimensional basins of attractions of
the system, which we have mapped out here for the case of a single short
optical perturbation. 

Finally, we highlight that the result presented here are quite
general in that they are generated only by excitability and delayed feedback. As
such, we believe that they will be of relevance beyond the scope of laser
dynamics for systems that encode information as pulse trains, e.g.,
those arising in neuroscience. 

$^{*}$These authors contributed equally to the work.

\bibliography{refs}

\end{document}